# Dynamic Superconductivity Responses in Photo-excited Optical Conductivity and Nernst Effect

Yasutomo J. Uemura

*Physics Department, Columbia University, New York, NY 10027, USA*

**Abstract**

**High-$T_c$ cuprate, alkali-doped $C_{60}$ and several other unconventional superconductors have very high transition temperatures $T_c$ with respect to the energy scale of superconducting (SC) charges inferred from the superfluid density (SFD). The observed linear relationship between $T_c$ and the SFD can hardly be expected in BCS superconductors while being reminiscent of Bose Einstein Condensation of pre-formed bosonic charges. As additional non-BCS like behaviors, responses similar to those in the bulk SC states have been observed at temperatures well above $T_c$ in the vortex-like Nernst effect, diamagnetic susceptibility, and transient optical conductivity in recent photo-excited pump-probe measurements. In this paper, we propose a coherent picture based on equilibrium and transient SFD to understand these unconventional behaviors in cuprates, $K_3C_{60}$, and organic superconductors. This picture assumes: (1) Dynamic SC responses in the Nernst and photo-induced measurements emerge at the formation of the local phase coherence (LPC) among wave functions of pre-formed bosonic pairs. (2) Its onset temperature $T_{LPC}$ is distinct from and lower than the boson formation temperature often denoted as the "pseudo-gap temperature" $T^*$, as $T_{LPC}$ is determined by the many-body boson density while $T^*$ represents attractive interaction between two fermions. (3) The bulk superconducting $T_c$, signaling global phase coherence, is significantly reduced from $T_{LPC}$, due to the competition between the SC and antiferromagnetic (AF) order. (4) The inelastic magnetic resonance mode (MRM) controls $T_c$ in the SC-AF competition. (5) The transient optical responses can be attributed to a change of the balance between the competing SC and AF orders caused by photo excitation. The assumptions (1) and (2) explain the relationship between $T_c$ and the transient SFD in photo excited studies and equilibrium SFD in Nernst effect. (3) and (4) are inferred from the linear dependence of $T_c$ on the MRM energy. (4) and (5) are consistent with the behaviors of the 400 cm$^{-1}$ optical responses in equilibrium and photo-excited studies and temperature dependence of the intensity of this optical mode and the MRM. Unlike previous phase-fluctuation pictures which expect dynamic responses between $T^*$ and $T_c$, the present picture involving competing order indicates that dynamic SC responses are seen between $T_{LPC}$ and $T_c$.**

## I. Introduction

Superconductivity is a dramatic manifestation of quantum physics, which was first discovered in Hg in 1911 [1]. More than 40 years after this in 1957, Bardeen, Cooper and Schrieffer presented BCS theory [2] which successfully explained superconductivity in simple metals, such as Sn, In, Pb, and Nb. In 1980's and afterwards, superconductivity has been discovered in heavy-fermion systems (1979) [3], organic metals (1980) [4,5], high-$T_c$ cuprates (1986) [6], alkali-doped $A_3C_{60}$ (1991) [7], and Fe-based systems (2006-2008) [8], in which extensive studies have revealed several features clearly different from those expected in BCS theory. These new systems are often referred to as "unconventional superconductors", but the exact mechanisms of their paring and condensation are yet to be established, even 30 years after the discoveries of cuprates [6].

The most notable "unconventional behaviors" in these systems include observation of the vortex-like Nernst effect [9,10] and diamagnetic susceptibility [11,12], pioneered by Ong and coworkers in cuprates in the 2000's, which exhibit responses reminiscent of bulk superconductors at temperatures well above $T_c$. Additional surprise was recently found in optical conductivity studies with photo-excitation by laser pumping pulses, performed by Cavalleri and co-workers in cuprates [13-18] and $K_3C_{60}$ [19,20], which revealed superconductor-like gapping and emergence of c-axis plasma edge in transient responses well above the equilibrium bulk $T_c$. Interpretation of these phenomena has, however, not yet been uniquely established: Nernst effect was ascribed primarily to SC-like vortices in [9,10] but to non-SC quasi particles and Fermi-surface effect in [21-23], and various models are being developed for the transient optical responses [24-27]. In the present paper, by examining relationship between their onset temperatures and the superfluid density (SFD), we propose a view that the vortex-like Nernst, diamagnetic and transient optical responses above $T_c$ emerge with the formation of local and dynamic phase coherence of pre-formed pairs.

## II. Phenomenological Plots

### (a) $T_c$ versus $T_F$ plot in equilibrium

Figure 1(a) shows a combination of a plot of the bulk $T_c$ versus effective Fermi temperature $T_F$ [28,29] together with several new points inferring different properties, denoted by the "photo-excited" and "Nernst onset" symbols, added for the present work. The horizontal axis $T_F$ represents the energy scale of SC charges derived from the Muon Spin Relaxation (MuSR) results of the SFD $n_s/m^*$ (SC carrier density / effective mass) at $T \rightarrow 0$ [30,31]. In BCS superconductors, $T_c$ is related to the energy gap, while there is no direct dependence of $T_c$ on the carrier density. This situation is illustrated in Fig. 1(b): BCS superconductors with the energy gaps shown by the black and red broken lines would have a factor of 2 different $T_c$ values, while their SFD will be identical because all the charge carriers (shown by the blue sphere) will join superfluid as long as any small gap is generated around the Fermi surface. There is an indirect dependency of $T_c$ on carrier density through the density of states (DOS) at the Fermi level in 3-dimensional (3-d) BCS systems. In ideally 2-d systems where the DOS does not depend on charge doping, however, even this indirect dependence vanishes. In contrast, Fig. 1(a) and MuSR results on cuprates [28,30] exhibit a strong dependence of $T_c$ on the SFD. This provided one of the earliest messages suggesting that condensation mechanism of cuprate superconductors may be distinctly different from that of BCS systems. Similar observations followed later in $A_3C_{60}$ [31] and Fe-based systems [29,32].

For given values of charge density $n_s$ and mass $m^*$, we can estimate the "hypothetical Bose Einstein Condensation (BEC) temperature" $T_{BEC} = T_F/4.6$, for the case if all the SC charges were forming a gas of tightly bound non-interacting pre-formed point-like bosons (each having charge 2e and mass 2$m^*$) from very high temperatures and there were no competing order. As illustrated in Fig. 1(c) for such an ideal Bose gas system, the spread of the wave function of each boson can be characterized by the thermal wave length $\lambda_{th}$, which becomes longer with decreasing temperature. BEC occurs when $\lambda_{th}$ becomes comparable to the inter-boson distance. Compared to $T_{BEC}$, the actual bulk $T_c$ of unconventional superconductors are reduced by at least a factor of 5. However, the linear relationship between $T_c$ and $T_F$, together with the universal and system-independent behavior for the upper-limit ratios of $T_c/T_F$ = 0.04~0.05 in Fig. 1(a), strongly suggests that the concept of BEC may be deeply related to condensation of these unconventional superconductors. Figure 1(a) represents an attempt to classify various systems between the limits of BEC and BCS-like behaviors. Doping dependence of cuprates has been discussed in terms of the BEC-BCS crossover picture [33,34].

**(b) Phase diagrams of cuprates with local phase coherence distinct from pair formation**

In order to sort out the relationship among various energy scales, we show the doping phase diagram of the (La,Sr)$_2$CuO$_4$ (LSCO) and YBa$_2$Cu$_3$O$_y$ (YBCO) cuprate systems in Fig. 2, together with the pseudogap temperature T* derived from the magnetic susceptibility and NMR [35,36], resistivity [35] and ARPES [37] measurements; onset-temperature of the vortex-like Nernst effect $T_{ner(v)}$ [10,38], quasi-particle Nernst effect $T_{ner(p)}$ [21-23] and diamagnetic susceptibility $T_{dia}$ [12,39]; bulk equilibrium transition temperature $T_c$; and the hypothetical BEC temperature $T_{BEC}$ derived from the observed SFD. The results of the LSCO system in Fig. 2(a) indicate that $T_{ner(v)}$ and $T_{dia}$ develop along with $T_{BEC}$ in the highly underdoped region.

Figure 2(b) shows a conceptual diagram with an assumption that T* represents the energy scale of gradual development of pair formation (e converted into 2e) with decreasing temperature. Then the normal state of the highly underdoped region will have mostly paired (2e) charges with few remaining unpaired fermion (e) charges. In this region, $T_{BEC}$ should represent the temperature where $\lambda_{th}$ becomes comparable to inter-(2e) distance, as illustrated in Fig. 1(c). The bulk $T_c$, characterized by the achievement of global phase coherence, however, is much lower than $T_{BEC}$, $T_{ner(v)}$ and $T_{dia}$ in this region, due to competing AF order as discussed later. Yet, the comparable length of thermal wavelength and the interboson distance would allow formation of at least local and dynamic phase coherence among wave functions of nearby bosons. Therefore, we term this temperature as the "local phase coherence temperature" $T_{LPC}$.

With increasing charge doping, T* decreases, and this makes existing boson number saturate and then decrease for a given T in the normal state. $T_{LPC}$, representing actual bosons (2e) co-existing with significant number of fermions (e), wildly departs from $T_{BEC}$ at a certain point, as shown in Fig. 2(b), and $T_{LPC}$ should be kept lower than T* since the local phase coherence develops only among paired 2e charges. If we assume that the vortex-like Nernst effect and diamagnetism require local phase coherence, $T_{ner(v)}$ and $T_{dia}$ will be represented by $T_{LPC}$. This argument provides an explanation for the peaking of $T_{ner(v)}$ and $T_{dia}$ at the Sr concentration of 0.10-0.12 and their location below T* in the LSCO system in Fig. 2(a). Similar tendency can be seen for YBCO in Fig. 2(c). The onset temperature $T_{ner(p)}$ of the quasi-particle Nernst effect [21-23], shown by the open blue diamond symbol, follows the behavior of T*, while $T_{ner(v)}$ of the vortex-like Nernst effect [10,38] and $T_{dia}$ [12,39] of diamagnetism follow the trajectory expected for $T_{LPC}$. The purple broken line in Fig. 2(c) indicates the onset temperature of the "precursor SC responses" found in equilibrium optical conductivity [40] at which the 320 cm$^{-1}$ phonon mode exhibits anomalies. This line also follows a trajectory expected for $T_{LPC}$.

In the transport response of highly anisotropic cuprate systems in the equilibrium state without photo-excitation, it has been known that the ab-plane conductivity increases below T* [41] while c-axis conductivity changes from metallic behavior above T* to semiconducting behavior below T* [42]. Naively, this can be understood in the scenario shown in Fig. 2(b). The pair formation below T*, which generates gapping in the excitation energy up to the attractive interaction between the paired two fermions, reduces the scattering rate in the ab-plane transport to which both e and (2e) charges would contribute. In contrast, the paired (2e) charges would have difficulty in moving towards the c-axis direction via tunneling, and this suppresses both the dc- and optical conductivity along the c-axis. At the SC transition, suppressed reflectivity of semiconducting c-axis optical responses above $T_c$ changes into a SC response with full reflectivity below $T_c$ accompanied by the "Josephson plasma edge" [36], which signals formation of coherent conduction along the c-axis.

**(c) Transient photo-induced phenomena in cuprates**

Since 2011, the research group of Cavalleri and co-workers have perturbed systems with photo excitation using a short pump laser pulse with energy corresponding to apical oxygen phonons, and observed (probed) transient responses in c-axis optical conductivity in several different cuprate systems [13-17,44]. In (La,Eu,Sr)$_2$CuO$_4$ (LESCO) [13,17] and (La,Ba)$_2$CuO$_4$ (LBCO) [18,41] with static spin stripe order and much reduced equilibrium bulk $T_c$, as well as in a very underdoped and non-SC YBCOy with y = 6.3, the c-axis plasma edge was absent in the equilibrium (non-SC) state at very low temperatures (T = 5-10 K), while photo-excitation induced emergence of plasma edge below T ~ 80 K in LESCO and LBCO, and at much higher temperatures in YBCO only in a short time scale after pumping. This suggests possible onset of photo-induced transient superconductivity-like behavior.

In optical conductivity σ, superconducting (SC) response can be estimated by the ω→0 limit of ω$Im$σ(ω). Comparing the equilibrium and transient values of this quantity, it has been shown that carriers for superfluid response have been generated in transient time for YBCO y=6.3. In Fig. 2(d), the results for ω$Im$σ(ω) are reproduced from ref. [15] for the y = 6.45, 6.5 and 6.6 SC samples of YBCO. Compared to the equilibrium superfluid response at low temperatures shown by the open circles, the photo excitation induced much more carriers in 6.45, comparable number in 6.5, while added much smaller SFD in 6.6 above $T_c$. For the y = 6.5 sample, we plot the onset temperature ~ 310 K of the superfluid response ω$Im$σ(ω) in Fig. 2(c) as $T_{photo}$ using the brown colored symbol at the corresponding chemical composition.

For the cuprates and other 2-d systems, the superfluid energy scales in Fig. 1(a) and $T_{BEC}$ in Fig. 2 (a) and (c) have been obtained from the ab-plane penetration depth. Unfortunately, it is not possible to estimate the transient SFD in the ab-plane optical responses in the above mentioned cuprates, since they are "too-good" ab-plane metals in the normal state and the onset of superconductivity makes only a very small change in the ab-plane reflectivity. In these YBCO systems, the ratios of the ab-plane vs c-axis values depend strongly on doping level both in dc-conductivity just above $T_c$ and in the equilibrium SFD at T→0. For YBCO y = 6.3 and 6.45 with photo-induced c-axis SFD far exceeding the equilibrium carrier density, it is not possible to estimate equilibrium doping level to which the photo-excited system corresponds. However, for

y = 6.5, the comparable values of the equilibrium and photo-induced c-axis SFD's at T→0 in Fig. 2(d) allow us to assume that the equilibrium and transient ab-plane SFD's are also likely comparable. With this assumption, we plot the onset temperature $T_{photo}$ of the transient SFD for y= 6.5 at the nominal equilibrium doping level in Fig. 2(c), and at the corresponding ab-plane equilibrium SFD and $T_F$ in Fig. 1(a).

If one plots $T_{photo}$ ~ 170 K of the y = 6.6 YBCO in Fig. 2(c) with the nominal chemical composition, the point lies close to $T_{dia}$. Both of the photo-induced SC and vortex-like Nernst / diamagnetic effects should selectively reflect responses of bosonic 2e carriers. The location of the point for y = 6.6, however, should be taken with a caution since the ab-plane response is unknown for the photo-excited state. The photo-induced c-axis superfluid response above $T_c$ quickly dies away with approach to the optimal doping region due to the decreasing population of pre-formed 2e pairs caused by the steep reduction of the pair formation temperature T*. Thus, most of the results of LSCO and YBCO in Fig. 2 can be explained with the scenario of Fig. 2(b).

**(d) Phase diagrams of BEDT and $A_3C_{60}$ systems with photo-excited and Nernst effects**

Similar tendency is seen in the Nernst effect results in the organic 2-d superconductors. Figure 3(a) shows the phase diagram against effective pressure, including $T_{BEC}$ derived from the MuSR SFD in (BEDT-TTF)$_2$-X with X = Cu(NCS)$_2$ (point D in Fig. 3(a)) and Cu[N(CN)$_2$]Br (point C) in ambient pressure [45] and from the magnetization results of Cu(NCS)$_2$ on the SFD in applied pressure [46]. It also shows the Nernst onset temperature $T_{ner}$ in X = Cu[N(CN)$_2$](Br,I) [47] for Br 100% (C), 80 % (B) and 73% (A) samples. Although nearly all the "underdoped" region is masked by the magnetic order, $T_{ner}$ lies close to $T_{BEC}$ near the boundary to the magnetic order, and the Nernst effect dies away rapidly with increasing (chemical) pressure.

As shown in Fig. 3(b), $A_3C_{60}$ (A = alkali metal Rb, K, Cs) systems also exhibit phase diagram, very similar to that of the 2-d organic superconductors [48]. Points for $T_{BEC}$ in this figure with pink-colored symbols are obtained by MuSR measurements of the equilibrium SFD in ambient pressure [31,49], and bulk $T_c$ from ref. [48]. Photo-induced transient optical responses have been published in K$_3$C$_{60}$ in ambient [19] and ambient plus applied pressure [20], with the latter study [20] performed using a higher laser power (fluence) than the former [19]. Their transient superfluid responses in $\omega Im\sigma(\omega)$ at the $\omega \to 0$ limit in ambient pressure are compared with the equilibrium spectral weight in Fig. 3(c).

For the low-fluence measurements, the low temperature photo-induced spectral weight is about a factor of 1.2 larger than the equilibrium value. In order to plot this result in Fig. 3(b), we scaled $T_{BEC}$ from the equilibrium MuSR SFD by this factor to obtain $T_{BEC}$ shown by the blue star symbol. In Fig. 3(b), we plot an estimate of $T_{photo}$ of the low-fluence measurements by the orange square symbol, to show that $T_{BEC}$ and $T_{photo}$ agree well. The high-fluence results [20] indicate that the photo-induced effect rapidly dies away with increasing hydrostatic pressure [20], and this makes the phase diagram of $A_3C_{60}$ system (Fig. 6 of ref. [20]) similar to that of the 2-d organic system (Fig. 3(a)). This similarity provides another support to our view that the Nernst and photo-excited responses emerge in the same area of the phase diagram and near $T_{BEC}$ before these effects quickly die away in the higher pressure region shown in Figs. 3(a) and (b). A similar tendency is seen in the higher doping regions of cupartes shown in Figs. 2(a) and (c).

**(e) Including dynamic superconductivity responses in the $T_c$ vs $T_F$ plot**

$A_3C_{60}$ is a cubic system, and the real part of the optical conductivity $Re\sigma(\omega)$ exhibits gapping of the Drude spectral weight below $T_c$ in equilibrium measurements [50,51]. The SFD can be estimated by the integrated area of this missing Drude weight which condenses as the delta function at $\omega = 0$. The gapping of the Drude spectral weight in $Re\ \sigma(\omega)$ in transient photo-induced responses can be seen for the low-fluence measurements [19] of $K_3C_{60}$ in ambient pressure. One finds a clear signature of gapping up to T = 100 K, but the response at 200 K is less clear in whether it is gapped or not. Therefore, we chose to make a crude estimate of $T_{photo}$ ~ 150 K for this case. To obtain the SFD for the low-fluence measurement at T→0, we scaled the MuSR SFD by multiplying 1.2 to the equilibrium MuSR value. We then obtained $T_F$ and plot $T_{photo}$ in Fig. 1(a) with a composite symbol surrounded by the open square.

In Fig. 1(a) we also plot $T_{photo}$ of YBCO 6.5 with the same composite symbol, and $T_{ner(v)}$ of LSCO (Sr 0.08), YBCO (y = 6.4) and $(BEDT-TTF)_2Cu[N(CN)_2](Br,I)$ with another composite symbol surrounded by the open circle. For the BEDT system with (Br,I) substitutions, the SFD results are available only for Br 100% denoted by C. For Br 80% (B) and 73% (A), we assumed the same SFD obtained in Br 100%, in view of little change of $T_{BEC}$ derived from the SFD in Fig. 3(a). The points for BEDT(A) and the above-mentioned other systems lie close to the $T_{BEC}$ line. $URu_2Si_2$ exhibits the onset of the Nernst effect [52] at the "hidden order transition temperature" $T_{hid}$ ~ 17 K, while the equilibrium $T_c$ ~ 1.3 K is much lower [53]. In Fig. 1(a), $T_{ner}$ of $URu_2Si_2$ also falls on the $T_{BEC}$ line. This might be explained if the pre-formed bosonic charges appear below $T_{hid}$, but the full understanding of the hidden order and verification of this guess require further studies [54]. The points of $T_{photo}$, $T_{ner(v)}$ and $T_{ner}$ in Fig. 1(a) indicate that these dynamic or transient responses reminiscent to bulk superconductors emerge at $T_{BEC}$ well above $T_c$. Hence a dynamic story is added to the previously static story of equilibrium state in Fig. 1(a) [28].

## III. Effects of competing order

**(a) Magnetic resonance mode as thermal excitations determining $T_c$**

We now consider why the actual $T_c$ is strongly reduced from $T_{BEC}$ and $T_{LPC}$ even if the local phase coherence emerges at $T_{BEC}$ in the underdoped systems. In BCS superconductors where all the charges in the normal state above $T_c$ are unpaired fermions, thermal process to determine $T_c$ is known to be the pair-breaking excitation across the energy gap. In systems shown in Figs. 2(a)-(c), however, the normal state of a wide doping region has dominant paired (2e) charges co-existing with unpaired fermion (e) charges. In this situation, thermal excitations should be a mixture of both the pair-breaking and pair-non-breaking processes, and their intensity ratios should reflect the population ratios of the unpaired (e) and paired (2e) normal-state charges just above $T_c$, which provide the final states of the thermal excitations.

The magnetic resonance mode (MRM) is a plausible candidate as the main thermal excitation process for destroying the SC condensate. The MRM has often been discussed as the pair-breaking excitation mode of d-wave (or s± wave) superconductors [55], assuming BCS-like situation with fully fermionic normal state. This is consistent with the SC gap energy scaling with the MRM energy [56]. For BEC-like situation, the present author proposed [34,57] a view that the MRM represents pair-non-breaking excitations of SC condensate with an analogy to rotons in superfluid $^4$He. Figure 4(a) shows correlations between the MRM energy and $T_c$ seen in various unconventional superconductors. For cuprates in which the MRM exhibits a complicated hour-glass shape dispersion, we plot the lower end energy of populated dispersion, which is often denoted as the spin-gap (SG) energy. The clear linear scaling indicates that the MRM energy is an important determining factor of $T_c$. This scaling is followed by many systems, including optimally doped cuprates with predominantly fermionic normal-state charges, highly underdoped cuprates with dominant pre-formed 2e pairs above $T_c$, as well as completely bosonic $^4$He which is unbreakable to fermions. This universality is likely coming from the dual character of the MRM as a mixture of pair-non-breaking and pair-breaking excitations.

The MRM also has a character as an inelastic excitation associated with the short range and dynamic AF (or stripe) spin fluctuations in the SC state related to the imminent competing spin order. This is similar to rotons having a character (among other He-specific features) as short-range and dynamic fluctuations of atomic configurations related to competing solid He state with the hexagonal closed packing (HCP) structure. Both excitations appear in the periodicity of these competing states, as illustrated in the inset figures of Fig. 4(a), and their mode energy controls the onset temperature of the adjacent SC or superfluid (SF) states [34,57]. When free energy of the SC (or SF) state is close to the competing AF (or HCP) state, competition of these states leads to the free energy landscape mimicking a double-well potential, as illustrated in Fig. 4(b). In this situation, the ordering temperature of the SC ground state is strongly influenced (reduced) by the existence of the competing spin states. The free energy difference between the ground and competing states is related to the closeness of these states (in doping or pressure tuning), and the inelastic excitation to the disordered state which controls the ordering temperature would contain mixed character of both SC and AF fluctuations. These features are consistent with the behavior of the MRM discussed above. The competition between the SC and AF phases with the free energy shown in Fig. 4(b) would result in first order transition at the boundary of these two phases. This is indeed the case for most of unconventional superconductors as pointed out in [57]. Thus, availability of the low-energy MRM excitations would reduce $T_c$ from $T_{BEC}$ and $T_{LPC}$ in the underdoped cuprates.

**(b) Photo excitation changing balance between the competing SC and AF order**

So far, we have seen that the photo-excitation causes difference between the equilibrium and photo-induced charge density at T = 0 in Figs. 2(d) and 3(c). This alone, however, would have caused much more modest change of the onset temperature of the SC response. If the photo excitation also temporarily changes the balance between the SC and AF order, we can expect a significant and temporal increase of $T_c$. For proceeding this argument, it would be desirable if the MRM response is observed in the transient time after photo excitation. Although transient neutron studies of the MRM have not yet been reported, suppression of the intensity of the "400 cm$^{-1}$ optical c-axis response" was observed in an underdoped YBCO 6.5 [14,16].

This is shown in Fig. 3(d) as the difference of optical conductivity between the photo-excited minus equilibrium responses. The spectral weight between 350 and 450 cm$^{-1}$ is reduced (black) while that of 280-350 cm$^{-1}$ is increased (red). The same shift of spectral weight has been observed in the c-axis equilibrium optical conductivity between T = 10 K and 300 K by Bernhardt and co-workers [40] as shown in Fig. 3(e). The increase of the 280-350 $cm^{-1}$ spectral weight was ascribed to increase of the intensity of 320 $cm^{-1}$ phonons, and the anomalous frequency shift of that phonon was discussed by these authors as a signature of "precursor superconductivity" [40]. The purple broken line of Fig. 2(c) shows the onset temperature of the phonon anomaly. The intensity of the 400 $cm^{-1}$ mode increases below the same temperature [40], as shown by the black "445 cm" symbol in Fig. 2(c).

The 400 $cm^{-1}$ mode has been attributed to the transverse Josephson plasma resonance [16,58,59] related to $CuO_2$ bilayers. In addition, Timusk and Homes [60] noted that its intensity exactly follows the temperature dependence of the MRM intensity, as shown in Fig. 3(f) [60,61]. If we adopt a view that the 400 $cm^{-1}$ mode intensity can represent the MRM as an indicator, then the transient loss of the intensity in Fig. 3(d) implies a transient removal of the MRM and suppression of the AF fluctuations leading to transient promotion of SC behavior. In YBCO 6.5, the photo excitation resulted in a very high transient $T_c$, as shown in Fig. 2(c) and (d).

The diamagnetic signal observed above $T_c$ is typically several orders of magnitude smaller than the Meissner signal of bulk superconductors below $T_c$ [12]. In contrast, in the pump-probe laser experiments, the transient response emerges with the signal intensity of at least more than 10% of that of equilibrium responses, which implies that significant portion of the charges in the optically-pumped volume participate in the transient SC responses. Although limited to a short transient time, superconductivity comparable to bulk superconductors is induced by the photo excitation. This can be expected only with a drastic change of controlling parameter(s), such as the above-mentioned SC/AF balance change.

In LBCO, the photo excitation is shown to "melt" stripe order, as evidenced by a temporal suppression of the Bragg peak intensities for charge order and LTT distortion [18,44]. while transient superfluid response appears in the c-axis optical conductivity. In the equilibrium state of LESCO, it has been noticed [17] that the charge stripe ordering at Sr content x = 0.125 suppresses superconductivity by disrupting the coherence along c-axis, while disappearance of charge order by increasing or decreasing x leads to restoration of the SC order. Photo excitation on the charge ordered LESCO was found to restore the transient c-axis SC responses [13,17]. These cases demonstrate that the destruction of competing static charge and spin order can promote superconductivity. The above-mentioned photo-induced perturbation on the MRM in YBCO implies that the transient suppression of the "dynamic mode" related to stripe spin-charge correlations may also promote superconductivity.

**IV. Discussions: relationship with earlier phase fluctuation pictures**

After the initial version of Fig. 1(a) was published in 1991 [28], an interpretation along the present view was discussed with a BEC-BCS picture by the present author [33,34], while Emery and Kivelson (EK) [62] discussed possible relevance to the Kosterlitz-Thouless (KT) transition in highly 2-d systems. Creation of vortex anti-vortex pairs in 2-d SC systems has been noticed in thin-film BCS superconductors [63] since the pre-high-$T_c$ era and has been discussed by Doniach and Huberman (DH) in terms of phase fluctuations of the SC wave function [64] above a KT transition. In both EK [62] and DH [64] pictures, however, the concept of competing order is absent, and the fluctuating SC phenomena is expected between T* and $T_c$ (which was not distinguished from $T_{LPC}$).

In contrast, in the 1991 paper [28], the present author and co-workers highlighted the effect of competing order by stressing the suppression of $T_c$ from $T_{BEC}$, and described the concept of phase fluctuations in the role of thermal wave length in BEC. Resorting to quantitative account for the onset temperature, the present work demonstrates that the dynamic SC response is observed between $T_{LPC}$ and $T_c$. It also introduces a new type of phase fluctuations, i.e., the pair-non-breaking thermal excitations of a SC pair into a (2e) pair in the normal state, as embodied in the MRM. Such phase fluctuations of bosonic charges can occur both in 2-d and 3-d systems. The similarities of isotropic 3-d $A_3C_{60}$ and weakly anisotropic FeAs to highly anisotropic cuprates and BEDT systems further support the present view. Absence of large single crystals has, however, so far prevented attempts of observing inelastic spin responses in $A_3C_{60}$ and BEDT systems.

## V. Summary

Since the present author and co-workers reported the correlations between $T_c$ and the SFD in 1989 [30] and $T_c$ and the effective Fermi temperature in 1991 [28], discussions on the SFD were confined mainly to the equilibrium physics of condensation at $T_c$ and the gap structure and pairing symmetry below $T_c$. In the present paper, we have newly extended consideration to the dynamic situation above $T_c$ by quantitative arguments based on the carrier density estimated from the SFD. Regarding the photo-excited studies, we introduced the concept of transient SFD and compared with the transient $T_c$ of cuprates and $A_3C_{60}$ systems in Figs. 1(a), 2(c) and 3(b). This has never been done to the knowledge of the present author.

Regarding the vortex-like Nernst effect and diamagnetism above $T_c$, the trajectory of the onset temperatures $T_{ner(v)}$, and $T_{dia}$ in the phase diagrams of cuprates has been well known, and has already been compared to signatures from other experiments such as anomalies in the optical studies (broken line of Fig. 2(c)) [40]. However, no successful attempts have been made to explain the magnitude of these onset temperatures. The phase-fluctuation models [62,64] have often been quoted as being consistent with the Nernst effect, despite the fact that the phase fluctuations set in below T* while the Nernst onset temperature is much lower than T* as shown in Figs. 2(a) and (c). The present work has shed a new light by comparing these onset temperatures to the hypothetical BEC temperature $T_{BEC}$ obtained from the SFD in Fig. 1(a), 2(a), 2(c), and 3(a) for cuprates, BEDT and $URu_2Si_2$.

These new plots revealed the following features: (1) there is a definite similarity between the photo-excited phenomena and the vortex-Nernst / diamagnetism effects. (2) in the highly underdoped side of the cuprates and near the phase boundary to antiferromagnetic phase in BEDT and $K_3C_{60}$ systems, the onset temperatures of both the photo-excited and Nernst / diamagnetism effects agree well with $T_{BEC}$. At this energy scale, one expects the formation of local phase coherence among pre-formed pairs. (3) Upon further doping in cuprates and chemical pressure in BEDT and $A_3C_{60}$, the onset temperature is rapidly reduced and merges with $T_c$. This behavior in the cuprates is consistent with the model shown in Fig. 2(b), where the normal state is characterized by co-existing bosons and fermions below T*, and the onset temperature is determined selectively by the pre-formed boson density.

The present work also elucidated the role of competing order in controlling $T_c$. In earlier papers [34,57], the present author proposed a model where the MRM plays a dominant role for thermal excitation in determining $T_c$. In the present paper, by assuming that MRM intensity can be represented by the 400 cm$^{-1}$ optical mode, we proposed a picture that the photo-excited superconductivity occurs due to a transient change of the balance between the SC and AF states, manifesting in transient removal of the MRM and the 400 cm$^{-1}$ mode. Among various theoretical proposals attempting to explain the photo-excited optical results, this is a new picture built upon observed transient evolution of the optical spectral weight and the above-mentioned phenomenology for the MRM. This picture is qualitatively consistent with the transient melting of magnetic order caused by photo excitation.

Due to highly successful BCS theory, researchers' mind is often locked exclusively into the BCS pictures, such as $T_c$ determined by the pair-breaking excitations across the energy gap, simultaneous occurrence of pair-formation and local/global phase coherence, pair-formation resulting from many-body physics, and requirement of low dimensionality for phase fluctuations. In the present work we have shown that these phenomena can develop in very different ways in unconventional superconductors, but the representative new observations, including Nernst, diamagnetic and photo-excited responses, can be understood in a coherent picture involving a clear distinction of the three processes, i.e., (1) pair (boson) formation at T*, (2) dynamic / local phase coherence at $T_{LPC}$, and (3) global SC transition winning against competing order at $T_c$.

**Acknowledgement**: The author would like to thank discussions with Andrea Cavalleri and Daniele Nicoletti, who kindly provided Figure 3(c) by re-plotting the data in their work published in refs. [19,20], and useful discussions with Dimitri Basov and Leon Degiorgi on optical responses. A part of the present paper was written when the author was visiting Paris as an Alliance Visiting Professor of Ecole Polytechnique in 2017-18 and as a CNRS Visiting Researcher of Sorbonne University in 2019. This work was supported by the US National Science Foundation grants DMR-1610633 and the DMREF project DMR-1436095, the Reimei Project from the Japan Atomic Energy Agency (JAEA), and a support from the Friends of Tokyo University Inc.

**Figure Captions**

Figure 1: *Plot of $T_c$, $T_{ner}$, $T_{photo}$ versus $T_F$, and illustration of the BCS and BEC condensation.* (a) The SC transition temperature $T_c$, the Nernst onset temperature $T_{ner(v)}$ [12,38] and $T_{ner}$ [47,52], plotted versus the effective Fermi temperature $T_F$ derived from the superfluid density $n_s/m*$ at the T = 0 limit obtained in MuSR measurements [28–31]; and the onset temperature of the photoexcited transient superconductivity $T_{photo}$ [15,19], plotted against $T_F$ derived from the transient superfluid density at the T=0 limit in the optical spectral weight. The broken line $T_{BEC}$ shows the Bose Einstein Condensation (BEC) temperature for non-interacting ideal Bose gas of density $n_s/2$ and mass $2m*$ expected in the absence of competing order. (b) Illustration of BCS condensation where all the SC carriers in the Fermi sphere, shown by the blue circle, participate in the superfluid, regardless of the size of the energy gap shown by black or red broken lines. (c) Illustration of BEC of an ideal boson gas in which condensation occurs when the thermal wave length $\lambda_{th}$ of a boson becomes comparable to the inter-boson distance.

Figure 2: *Phase diagrams of high-$T_c$ cuprates, and photo-induced superconducting responses.* Phase diagrams of (a) $(La,Sr)_2CuO_4$ (LSCO) and (c) $YBa_2Cu_3O_y$ (YBCO) high-$T_c$ cuprate systems including the pseudo gap temperature T* from magnetic susceptibility (chi) [35], dc conductivity (rho) [35], NMR [36] and ARPES pseudo gap [37], $T_{BEC}$ derived from the T=0 superfluid density $n_s/m*$ measured by MuSR [26], the onset temperatures of the vortex-like Nernst effect $T_{ner(v)}$ [12,38], quasi-particle Nernst effect $T_{ner(p)}$ [21-23] and diamagnetic susceptibility $T_{dia}$ [12,39], equilibrium superconducting $T_c$ [12,15] and the onset temperature $T_{photo}$ of the photo-induced transient superconducting responses [15]. In (c), the results from the equilibrium c-axis optical conductivity [40] are also plotted: the purple broken line denotes the onset of "precursor superconductivity" derived from the 320 cm$^{-1}$ phonon anomaly; the black symbol shows the onset temperature of the transverse Josephson mode intensity around 445 cm$^{-1}$, and the green "phonon" symbol denotes T* derived from the spectral weight between 140 and 660 cm$^{-1}$. (b) shows a conceptual phase diagram for a model which assumes T* as the pair formation temperature. $T_{LPC}$ represents the onset temperature of the local phase coherence among pre-formed pairs. (d) Transient superconducting responses corresponding to the superfluid density, obtained from the imaginary part of the optical conductivity observed in photo-excited measurements in YBCO systems [15]. (reproduced with the permission of the authors).

Figure 3: *Phase diagrams of BEDT and $A_3C_{60}$ superconductors, photo-excited responses, transient and equilibrium spectral weight changes and comparison with MRM intensity in YBCO*
Phase diagrams of (a) organic (BEDT-TTF)$_2$-X systems with X = Cu(NCS)$_2$ and Cu[N(CN)$_2$](Br,I) and (b) $A_3C_{60}$, both as a function of effective chemical / hydrostatic pressure represented by the lattice constant. Included are the superconducting $T_c$, [46-48], Nernst effect onset temperature $T_{ner}$ [47] for BEDT with Br 73% (A), 80% (B) and 100% (C), $T_{BEC}$ from the superfluid density estimated by MuSR [31,45] and magnetization [46] in equilibrium and by transient photo-induced optical conductivity [19], antiferromagnetic ordering temperature $T_{AF}$ [47,48], and the onset temperature of photo-induced superconducting responses in the low-fluence measurements of $K_3C_{60}$ [19]. (c) shows the temperature dependence of the superfluid density in $K_3C_{60}$ at ambient pressure derived from the imaginary part of the optical conductivity in photo-excited measurements with low fluence [19] and high fluence [20] laser intensity, compared with equilibrium measurement, kindly provided by the authors of [19] and [20]. (d) temporal change of the spectral weight of the c-axis optical conductivity $\sigma(\omega)$ of YBCO 6.5 [14] ($T_c$ = 50 K) caused by the photo excitation. (e) equilibrium $\sigma(\omega)$ in the same frequency range in YBCO 6.6 [40] ($T_c$ = 58 K). (f) comparison of the neutron MRM intensity in YBCO 6.6 ($T_c$ = 59 K) [61] and the 400 cm$^{-1}$ mode in $\sigma(\omega)$ depicted from Timusk and Homes [60]. Printed with copyright permissions from Springer-Nature for (d) and Elsevier for (f).

Figure 4: *Plot of $T_c$ versus resonance-mode / roton energies and the free-energy landscape.*
(a) Correlations between the superconducting / superfluid $T_c$ and the energy of the magnetic resonance mode (MRM) in various unconventional superconductors [56,57] and the roton-minimum energy in superfluid $^4$He [34,57] in several different pressures. For the YBCO and LSCO cuprates which exhibit the hour-glass shape dispersion of the MRM, we plot the lowest populated energy, often called as "spin gap" (SG) energy. Inset figures show MRM and roton excitations in the energy-momentum plane, together with the Bragg peak momentum for the competing spin-density-wave (SDW) order in a 122 FeAs superconductor [65] and hexagonal-closed-pack (HCP) order of solid He [34]. (b) illustrates a free energy landscape of superconducting (SC) state and competing antiferromagnetic (AF) state in unconventional superconductors. This landscape is consistent with first-order transition between SC and AF phases observed in many unconventional superconductors [57].

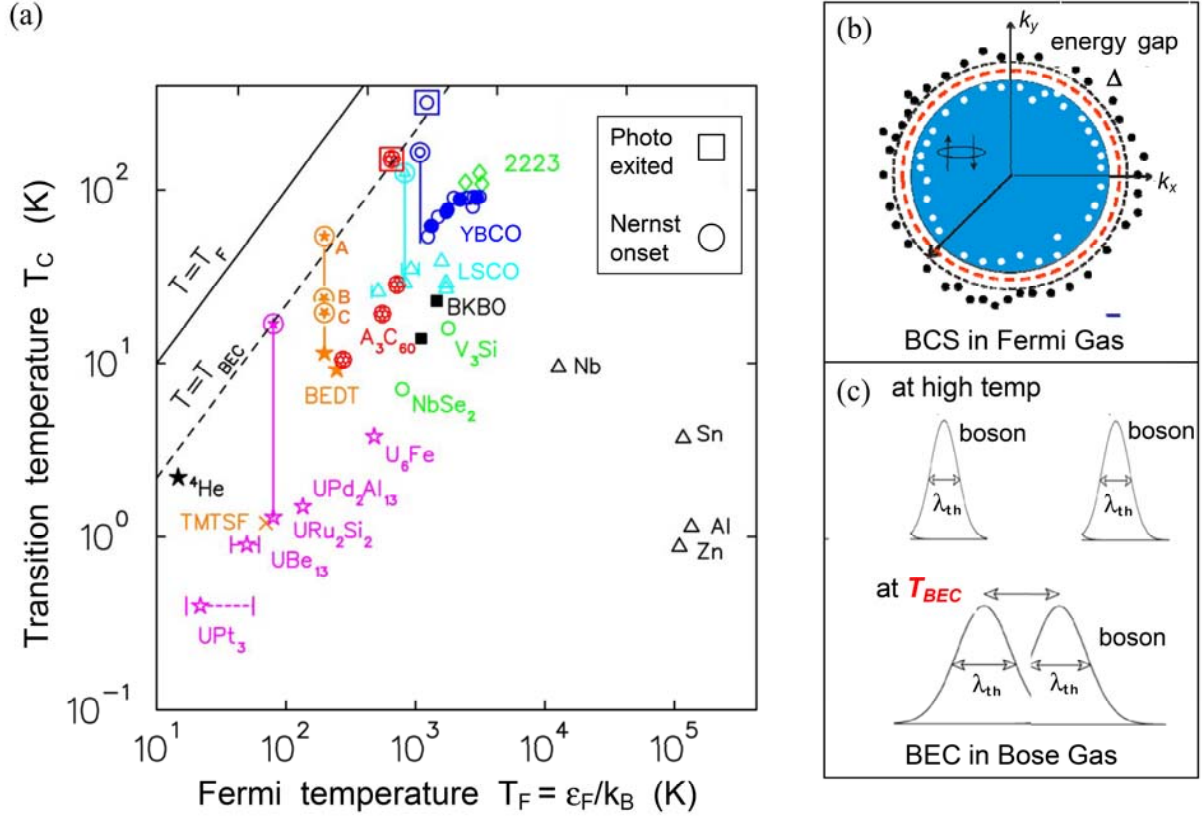

Figure 1: ***Plot of $T_c$, $T_{ner}$, $T_{photo}$ versus $T_F$, and illustration of the BCS and BEC condensation.***
(a) The SC transition temperature $T_c$, the Nernst onset temperature $T_{ner(v)}$ [12,38] and $T_{ner}$ [47,52], plotted versus the effective Fermi temperature $T_F$ derived from the superfluid density $n_s/m*$ at the T = 0 limit obtained in MuSR measurements [28–31]; and the onset temperature of the photoexcited transient superconductivity Tphoto [15,19], plotted against $T_F$ derived from the transient superfluid density at the T=0 limit in the optical spectral weight. The broken line $T_{BEC}$ shows the Bose Einstein Condensation (BEC) temperature for non-interacting ideal Bose gas of density $n_s/2$ and mass $2m*$ expected in the absence of competing order. (b) Illustration of BCS condensation where all the SC carriers in the Fermi sphere, shown by the blue circle, participate in the superfluid, regardless of the size of the energy gap shown by black or red broken lines. (c) Illustration of BEC of an ideal boson gas in which condensation occurs when the thermal wave length $\lambda_{th}$ of a boson becomes comparable to the inter-boson distance.

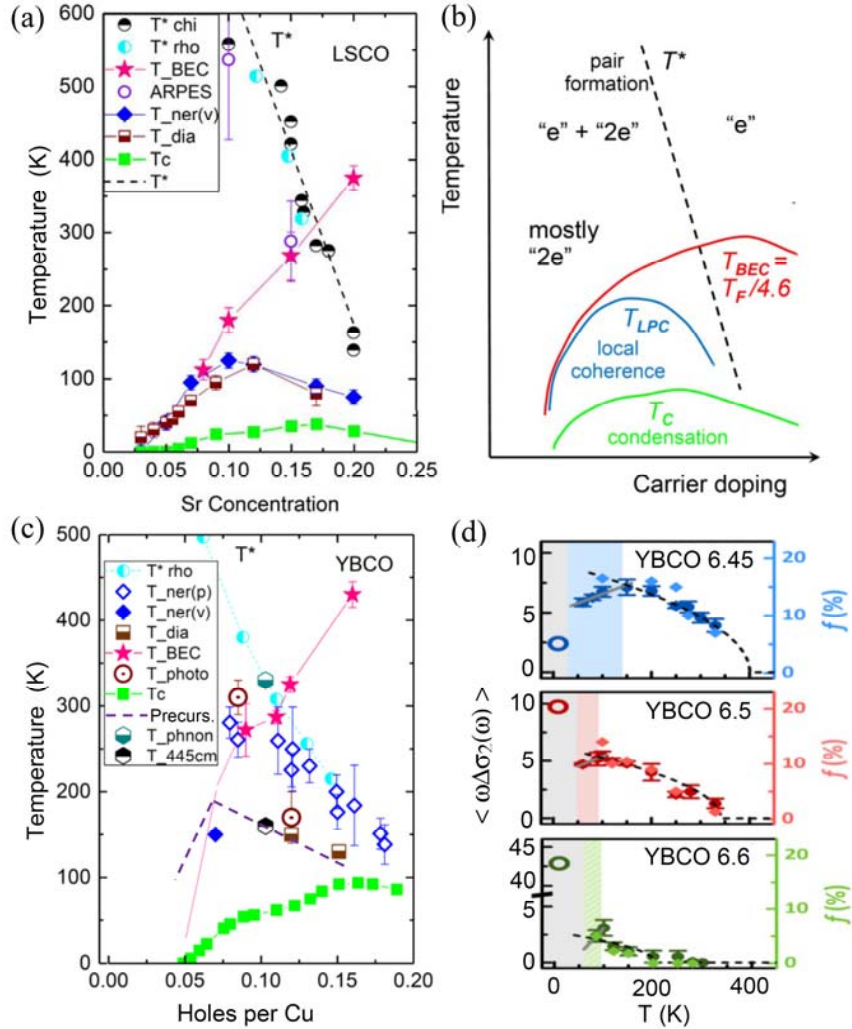

Figure 2: *Phase diagrams of high-$T_c$ cuprates, and photo-induced superconducting responses.*
Phase diagrams of (a) $(La,Sr)_2CuO_4$ (LSCO) and (c) $YBa_2Cu_3O_y$ (YBCO) high-$T_c$ cuprate systems including the pseudo gap temperature T* from magnetic susceptibility (chi) [35], dc conductivity (rho) [35], NMR [36] and ARPES pseudo gap [37], $T_{BEC}$ derived from the T=0 superfluid density $n_s/m^*$ measured by MuSR [26], the onset temperatures of the vortex-like Nernst effect $T_{ner(v)}$ [12,38], quasi-particle Nernst effect $T_{ner(p)}$ [21-23] and diamagnetic susceptibility $T_{dia}$ [12,39], equilibrium superconducting $T_c$ [12,15] and the onset temperature $T_{photo}$ of the photo-induced transient superconducting responses [15]. In (c), the results from the equilibrium c-axis optical conductivity [40] are also plotted: the purple broken line denotes the onset of "precursor superconductivity" derived from the 320 cm$^{-1}$ phonon anomaly; the black symbol shows the onset temperature of the transverse Josephson mode intensity around 445 cm$^{-1}$, and the green "phonon" symbol denotes T* derived from the spectral weight between 140 and 660 cm$^{-1}$. (b) shows a conceptual phase diagram for a model which assumes T* as the pair formation temperature. $T_{LPC}$ represents the onset temperature of the local phase coherence among pre-formed pairs. (d) Transient superconducting responses corresponding to the superfluid density, obtained from the imaginary part of the optical conductivity observed in photo-excited measurements in YBCO systems [15] (reproduced with the permission of the authors).

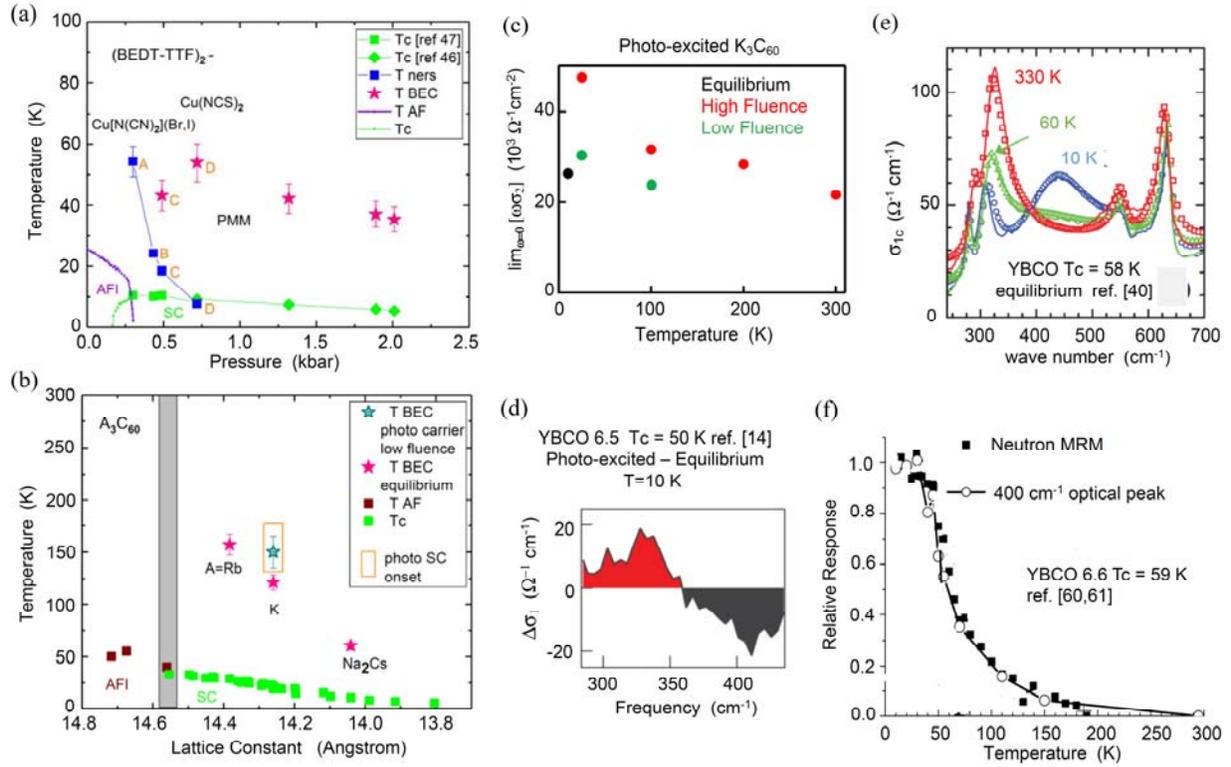

Figure 3: ***Phase diagrams of BEDT and $A_3C_{60}$ superconductors, photo-excited responses, transient and equilibrium spectral weight changes and comparison with MRM intensity in YBCO***
Phase diagrams of (a) organic $(BEDT-TTF)_2$-X systems with X = $Cu(NCS)_2$ and $Cu[N(CN)_2](Br,I)$ and (b) $A_3C_{60}$, both as a function of effective chemical / hydrostatic pressure represented by the lattice constant. Included are the superconducting $T_c$, [46-48], Nernst effect onset temperature $T_{ner}$ [47] for BEDT with Br 73% (A), 80% (B) and 100% (C), $T_{BEC}$ from the superfluid density estimated by MuSR [31,45] and magnetization [46] in equilibrium and by transient photo-induced optical conductivity [19], antiferromagnetic ordering temperature $T_{AF}$ [47,48], and the onset temperature of photo-induced superconducting responses in the low-fluence measurements of $K_3C_{60}$ [19]. (c) shows the temperature dependence of the superfluid density in $K_3C_{60}$ at ambient pressure derived from the imaginary part of the optical conductivity in photo-excited measurements with low fluence [19] and high fluence [20] laser intensity, compared with equilibrium measurement, kindly provided by the authors of [19] and [20]. (d) temporal change of the spectral weight of the c-axis optical conductivity $\sigma(\omega)$ of YBCO 6.5 [14] ($T_c$ = 50 K) caused by the photo excitation. (e) equilibrium $\sigma(\omega)$ in the same frequency range in YBCO 6.6 [40] ($T_c$ = 58 K). (f) comparison of the neutron MRM intensity in YBCO 6.6 ($T_c$ = 59 K) [61] and the 400 cm$^{-1}$ mode in $\sigma(\omega)$ depicted from Timusk and Homes [60]. Printed with copyright permissions from Springer-Nature for (d) and Elsevier for (f).

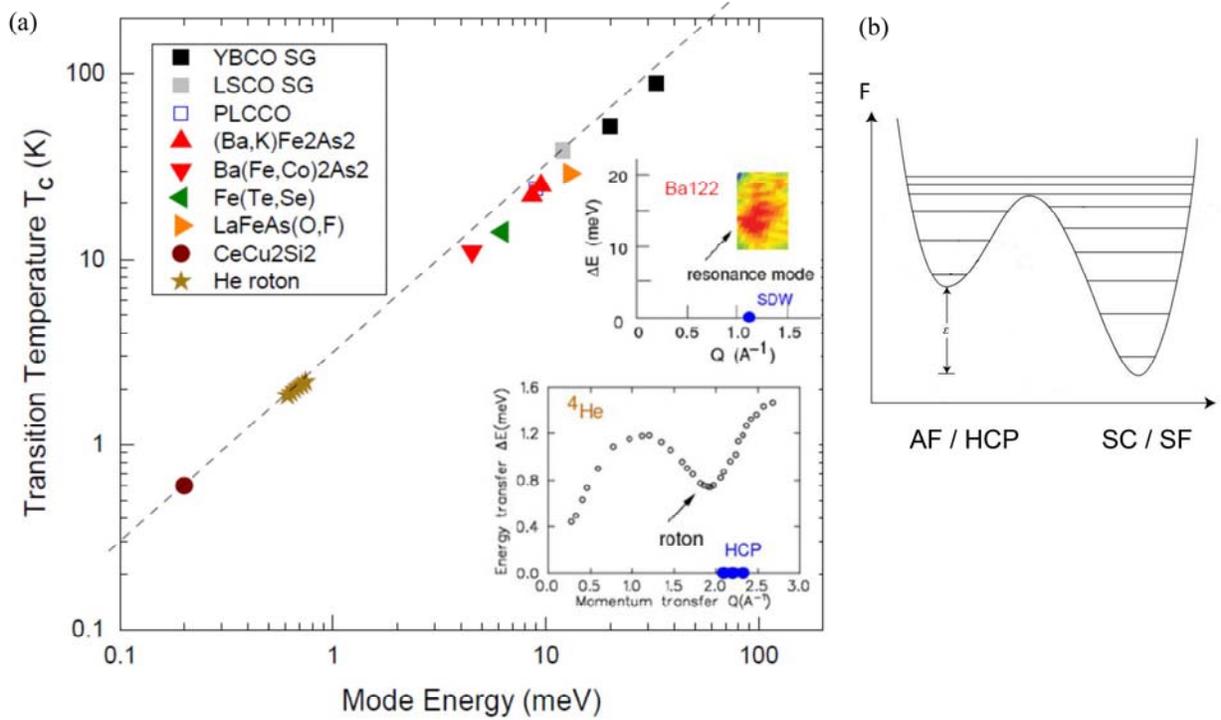

Figure 4: *Plot of $T_c$ versus resonance-mode / roton energies and the free-energy landscape.*
(a) Correlations between the superconducting / superfluid $T_c$ and the energy of the magnetic resonance mode (MRM) in various unconventional superconductors [56,57] and the roton-minimum energy in superfluid $^4$He [34,57] in several different pressures. For the YBCO and LSCO cuprates which exhibit the hour-glass shape dispersion of the MRM, we plot the lowest populated energy, often called as "spin gap" (SG) energy. Inset figures show MRM and roton excitations in the energy-momentum plane, together with the Bragg peak momentum for the competing spin-density-wave (SDW) order in a 122 FeAs superconductor [64] and hexagonal-closed-pack (HCP) order of solid He [34]. (b) illustrates a free energy landscape of superconducting (SC) state and competing antiferromagnetic (AF) state in unconventional superconductors. This landscape is consistent with first-order transition between SC and AF phases observed in many unconventional superconductors [57].